\documentclass{article}
\usepackage{caption}
\usepackage[numbers]{natbib}
\usepackage{graphicx}
\usepackage{placeins}
\usepackage{multirow}
\usepackage{url}
\usepackage{breakurl}
\usepackage{hyperref}
\hypersetup{breaklinks=true}

\usepackage[preprint]{neurips_2024}

\usepackage[utf8]{inputenc} 
\usepackage[T1]{fontenc}    
\usepackage{hyperref}       
\usepackage{url}            
\usepackage{booktabs}       
\usepackage{amsfonts}       
\usepackage{nicefrac}       
\usepackage{microtype}      
\usepackage{xcolor}         

\title{AI Meets Antimatter: \\ 
Unveiling Antihydrogen Annihilations}

\author{
      Ashley Ferreira \\
      University of Waterloo\footnotemark[1] \\ 
      Waterloo, ON, Canada \\
      \texttt{a4ferrei@uwaterloo.ca} \\
      \And
      Mahip Singh \\
      University of Waterloo\footnotemark[1] \\
      Waterloo, ON, Canada \\
      \texttt{mahip.singh@uwaterloo.ca} \\
      \And
      Andrea Capra \\
      TRIUMF \\
      Vancouver, BC, Canada \\
      \texttt{acapra@triumf.ca} \\
      \And 
      Ina Carli \\
      TRIUMF \\
      Vancouver, BC, Canada \\
      \texttt{icarli@triumf.ca} \\
      \And 
      Daniel Duque Quiceno \\
      TRIUMF \\
      Vancouver, BC, Canada \\
      \texttt{dduque@triumf.ca} \\
      \And 
      Wojciech T. Fedorko \\
      TRIUMF \\
      Vancouver, BC, Canada \\
      \texttt{wfedorko@triumf.ca} \\
      \And 
      Makoto M. Fujiwara \\
      TRIUMF \\
      Vancouver, BC, Canada \\
      \texttt{fujiwara@triumf.ca} \\
      \And 
      Muyan Li \\
      University of Waterloo\footnotemark[1] \\
      Waterloo, ON, Canada \\
      \texttt{a5li@uwaterloo.ca} \\
      \And 
      Lars Martin \\
      TRIUMF \\
      Vancouver, BC, Canada \\
      \texttt{lmartin@triumf.ca} \\
      \And 
      Yukiya Saito \\
      TRIUMF \\
      Vancouver, BC, Canada \\
      \texttt{ysaito@alum.ubc.ca} \\
      \And 
      Gareth Smith \\
      TRIUMF \\
      Vancouver, BC, Canada \\
      \texttt{gsmith@triumf.ca} \\
      \And 
      Anqi Xu \\
      University of British Columbia\footnotemark[1] \\
      Vancouver, BC, Canada \\
      \texttt{anqixuu@student.ubc.ca} \\
}

\begin{document}

\maketitle
\footnotetext[1]{
This research was conducted during the authors' affiliation with TRIUMF and their respective institutions.}

\begin{abstract}
The ALPHA-g experiment at CERN aims to perform the first-ever direct measurement of the effect of gravity on antimatter, determining its weight to within 1\% precision. This measurement requires an accurate prediction of the vertical position of 
annihilations within the detector. In this work, we present a novel approach to annihilation position reconstruction using an ensemble of models based on the PointNet deep learning architecture. The newly developed model, PointNet Ensemble for Annihilation Reconstruction (PEAR) outperforms the standard approach to annihilation position reconstruction, providing more than twice the resolution while maintaining a similarly low bias. This work may also offer insights for similar efforts applying deep learning to experiments that require high resolution and low bias.
\end{abstract}

\section{Introduction}
The ALPHA-g experiment is part of the ALPHA antihydrogen ($\overline{\mathrm{H}}$) program at CERN, with a goal of measuring the gravitational acceleration of $\overline{\mathrm{H}}$ to a precision of 1\% \cite{Chukman2020, Capra2017} as a step towards 
understanding the matter-antimatter asymmetry in the universe \cite{CERN2021}. In September 2023, the collaboration published the first measurement of the terrestrial gravitational acceleration of antimatter, determining that it does indeed fall downwards, but the measurement was far from the 1\% precision threshold \cite{anderson2023observation}.

In the ALPHA-g apparatus, $\overline{\mathrm{H}}$ atoms are held at a known location in a magnetic trap within a vacuum chamber then, when the magnetic trap is shut down, the $\overline{\mathrm{H}}$ atoms are released. When $\overline{\mathrm{H}}$ hits the sides of the vacuum chamber, it annihilates with ordinary matter. By recording the vertical position of annihilation and the time it occurred during magnet ramp-down (and thus the magnetic field), the effect of gravity can be calculated. 

The spatial position of the annihilations can be inferred from the charged particles produced in the annihilation process \cite{Capra2017}. The interaction of the charged particles in the radial Time Projection Chamber (rTPC) produces electrical signals, which are combined to form ``spacepoints'' by a suitable software. Each spacepoint is a triplet of Cartesian coordinates that represents the location of the interaction in the active volume of the detector. The current method for predicting the annihilation positions, called ``Helix Fit,'' begins by grouping these spacepoints into clusters and fitting each cluster with a 3D helix function. Then, an annihilation position is found at the location where at least two helices pass closest to each other, which is termed the ``vertex'' \cite{anderson2023observation}.

\begin{figure}[htbp]
  \centering
  \includegraphics[width=0.85\textwidth]{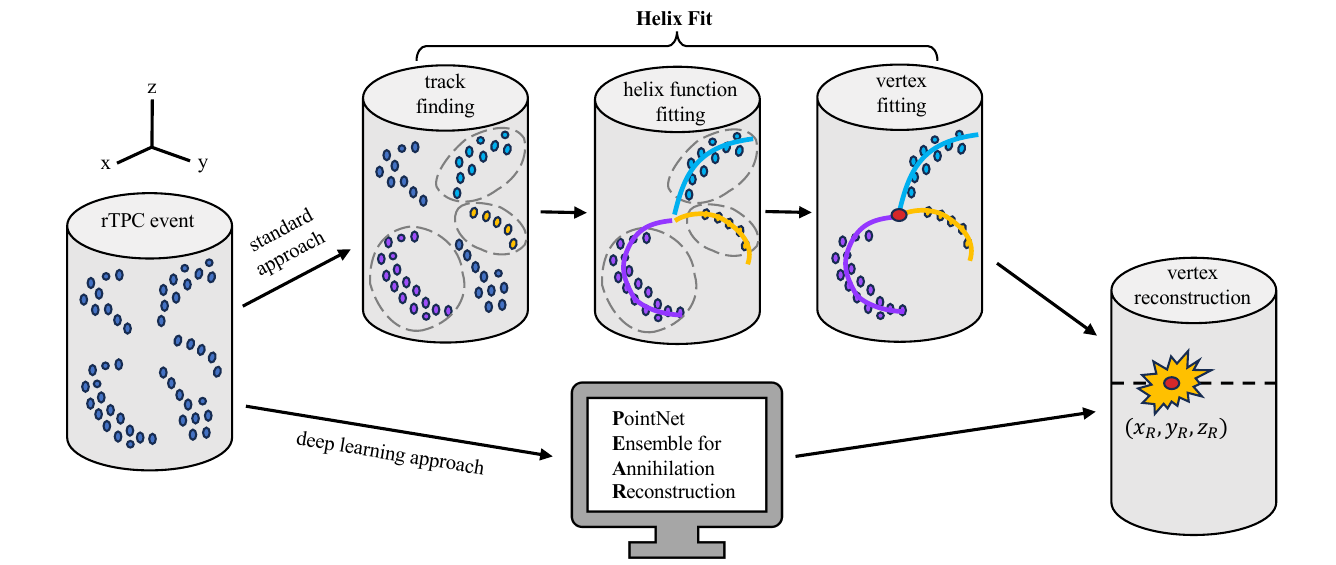}
  \caption{\label{fig:concept}Conceptual schematics of the vertex reconstruction approach using our deep learning model (bottom), in contrast to the standard method that requires identification of particle tracks and fitting helix functions (top).}
\end{figure}

In this work, we introduce the PointNet Ensemble for Annihilation Reconstruction (PEAR), which directly predicts the vertical position of the $\overline{\mathrm{H}}$ annihilation for the ALPHA-g experiment from spacepoints using deep learning. This bypasses the challenging intermediate steps required by Helix Fit, such as track identification and fit optimization procedures. Our new approach takes in all the information available from the detectors, while the conventional Helix Fit discards spacepoints that are not identified as specific types of particles. The conceptual schematics of this approach, along with the Helix Fit procedure, is shown in Figure~\ref{fig:concept}. Similar machine learning techniques have been applied in other contexts involving detectors like the Active-Target Time Projection Chamber (AT-TPC), where they were used for nuclear track classification in low-energy nuclear physics experiments studying exotic nuclei \cite{Kuchera2019}.

\section{Methods}

Our method uses a modified version of the deep learning model architecture called PointNet \cite{Qi2017}. PointNet takes inputs of spacepoints directly as a 3D point cloud, avoiding potential information loss or added computational overhead compared to architectures that require converting point cloud data into structured formats like 2D grids or voxel representations \cite{Qi2017}. 
PointNet was originally developed for the segmentation and classification of 3D point cloud data. To use this model for the prediction of annihilation positions, which is a regression task, the final fully connected layer has been replaced by a linear output and the input transformation layer has been removed.

\begin{figure}[htbp]
\centering
\includegraphics[width=\textwidth]{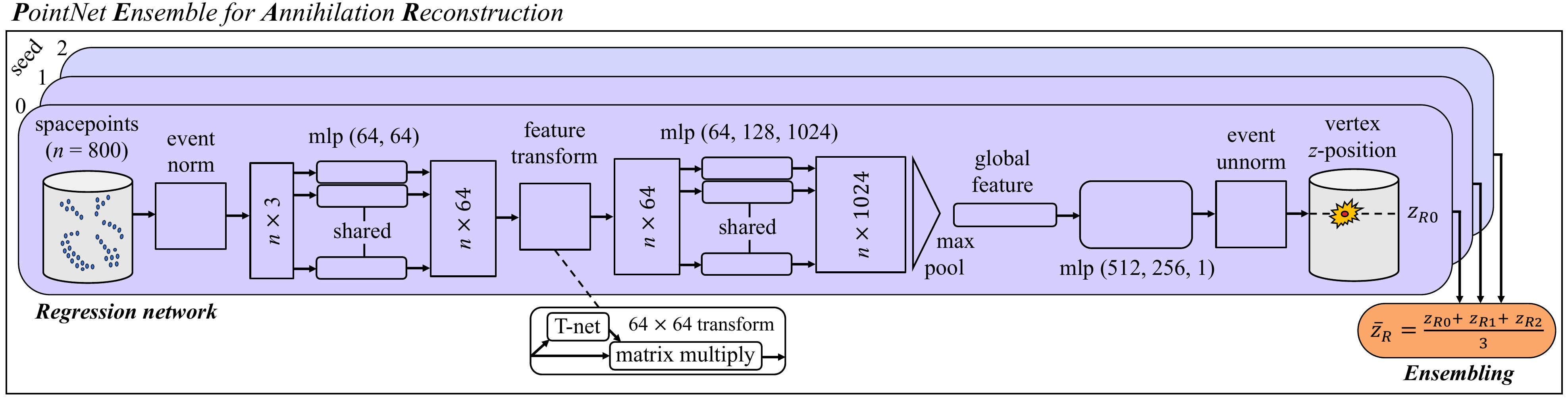}
\caption{\label{fig:architecture} Schematics of our modified PointNet architecture for the vertex reconstruction regression task, heavily based on \cite{Qi2017}.} 
\end{figure}

This model was trained using spacepoints from Monte Carlo simulations of the rTPC. A dataset of 2.7 million events was generated, where each event corresponds to one $\overline{\mathrm{H}}$ annihilation. Spacepoints are used as model inputs, and the target output value is the true $z$ coordinate of the vertex. This dataset was split into approximately 80\%  for training, 10\% for validation, and 10\% for testing. 

During training, we tracked Huber Loss and performance across additional metrics on half of the validation dataset each epoch. In particular, we introduced a new $z$-bias measure called the Absolute Residual Average (ARA) and saved checkpoints when the model exhibited ARA lower than Helix Fit.  To calculate the ARA, the detector is divided into 100\,mm slices along the $z$-axis, allowing us to examine the performance of the model at different positions along the detector. For each slice ($i = 1, \ldots, N_\mathrm{slices}$ where $N_\mathrm{slices}$ is the number of slices), we compute the residuals ($\Delta z_i = z_{\mathrm{Reconstructed}\,i} - z_{\mathrm{Simulated} \,i}$) and the average of these residuals ($\mu_i = \overline{\Delta z_i}$), followed by absolute averaging of these per-slice averages, as is shown in equation ~\ref{eq:overall_bias}.

\begin{equation}
\label{eq:overall_bias}
\mathrm{ARA} = \frac{1}{N_{\mathrm{slices}}}\sum_{i}^{N_{\mathrm{slices}}} |\mu_i|.
\end{equation}

From the checkpoints with low ARA recorded on the first half of the validation set, the one with the lowest ARA is selected as this yielded models with low loss while prioritizing minimal $z$-bias. The second half of the validation set was then used to confirm that the observed performance was not an artifact of checkpoint selection. This splitting of the validation set into two parts allowed us to reserve the test set solely for generating our final published evaluations. We followed this procedure to train three models with identical architectures, each initialized with a different random seed. We averaged the predictions of these three models to create an ensemble \cite{Breiman1996BaggingP}. In this case, an ensemble of three models (each the top epoch of their respective training run), minimized ARA, reduced loss, and improved the robustness of the model to changes in the data.

We normalized the inputs by subtracting the mean $z$ spacepoint value from all spacepoints $z$-coordinates, as well as from the true $z$-vertex for each event. The mean $z$ spacepoint value can then be added to the normalized $z$-vertex predicted by the model to get the unnormalized value for use in the final output. This led to significant improvements in resolution, defined as the ability to precisely determine the true $z$-vertex, as well as training efficiency. This is likely due to both the smaller numerical range that the model has to span and the events being approximately invariant to vertical translations, allowing the model to focus on learning the relationship between the relative position of the spacepoints and the corresponding $z$-vertex. By eliminating this redundant degree of freedom, the model focuses on a less complex representation of the data, effectively increasing the density of meaningful examples per unit of variation in the training space. This normalization method also proved highly effective in reducing $z$-bias by not providing the model access to information that would be undesirable to influence the model's outputs, in this case, translation along the $z$-axis. 

\section{Results}
\begin{figure}[htbp]
\centering
\includegraphics[width=0.9\textwidth]{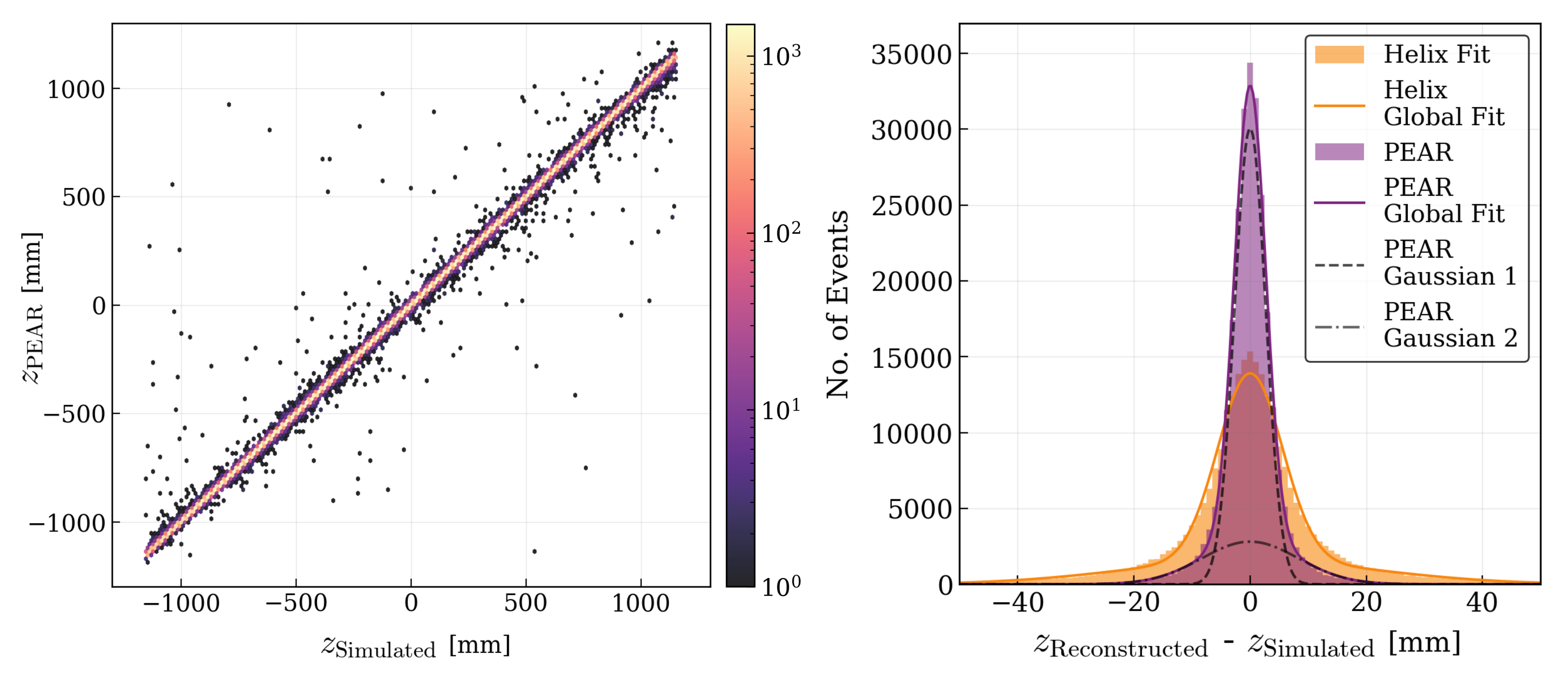} 
\caption{\label{fig:res_dist} Heat plot of predicted $z$-vertex with PEAR versus true $z$-vertex (left). Histogram of residuals for PEAR and Helix Fit with Gaussian fits (right).}
\end{figure}

Evaluating PEAR on the test dataset and comparing its performance to Helix Fit, we see that it achieves similar $z$-bias and significantly better resolution. To begin with, Figure~\ref{fig:res_dist} (left) shows the reconstructed $z$-vertex from PEAR's predictions versus the simulated $z$-vertex. There is a strong agreement between the predicted and true values, evidenced by a Pearson correlation coefficient of 0.9997.

\vspace{-2pt} 
 Figure~\ref{fig:res_dist} (right) displays the distributions of the residuals for PEAR and Helix Fit. The narrower peak around zero from PEAR indicates that it provides a higher resolution than Helix Fit, meaning that, on average, the $z$-vertex reconstruction from PEAR is closer to the true $z$-vertex compared to Helix Fit. 
 
 Table~\ref{tab:res_results} lists the summary statistics for the performance of both methods and shows that PEAR has better overall performance with a Full Width at Half Maximum (FWHM) less than half that of Helix Fit while maintaining a statistically consistent ARA, and consistently lower standard deviations ($\sigma$) on the fits applied. 
 We fit two Gaussians with shared means to represent the residual distribution, one for the core and one for the tail. We find that the $\sigma$ of each of the constituent Gaussians are smaller than the corresponding ones for Helix Fit. Moreover, the integrals show us that more of the distribution is represented by the core Gaussian for PEAR compared to Helix Fit, indicating that PEAR has residuals that are more tightly concentrated around the mean and has fewer outliers.

\begin{table}[h!]
\centering
\begin{tabular}{|l|l|l|l|}
\hline & Metric & 
\textbf{PEAR} & \textbf{Helix Fit} \\ \hline
\multirow{2}{*}{Gaussian 1} & $\sigma_1$ [mm] & $2.65$ $\pm$ $0.01$ & 5.52 $\pm$ 0.02\\ \cline{2-4} 
& Integral & 76.9\%& 66.3\%\\ \hline
\multirow{2}{*}{Gaussian 2} & $\sigma_2$ [mm] & $8.17$ $\pm$ $0.07$& $21.17$ $\pm$ $0.13$\\ \cline{2-4} 
& Integral & 22.5\%& 32.9\%\\ \hline
\multirow{2}{*}{Overall}
& FWHM [mm] & $6.62 \pm 0.02$ & $14.12 \pm 0.06$\\ \cline{2-4} 
& ARA [mm] & $0.04 \pm 0.01 $& $0.06 \pm 0.02$\\ \cline{2-4} 
\hline
\end{tabular}
\caption{Model performance comparison between PEAR and Helix Fit. Note the ARA calculation is at 100\,mm slices and there is a constant component to the fit accounting for <1\% of the integral.
}
\label{tab:res_results}
\end{table}

\begin{figure}[htbp]
\centering 
\includegraphics[width=0.75\textwidth]{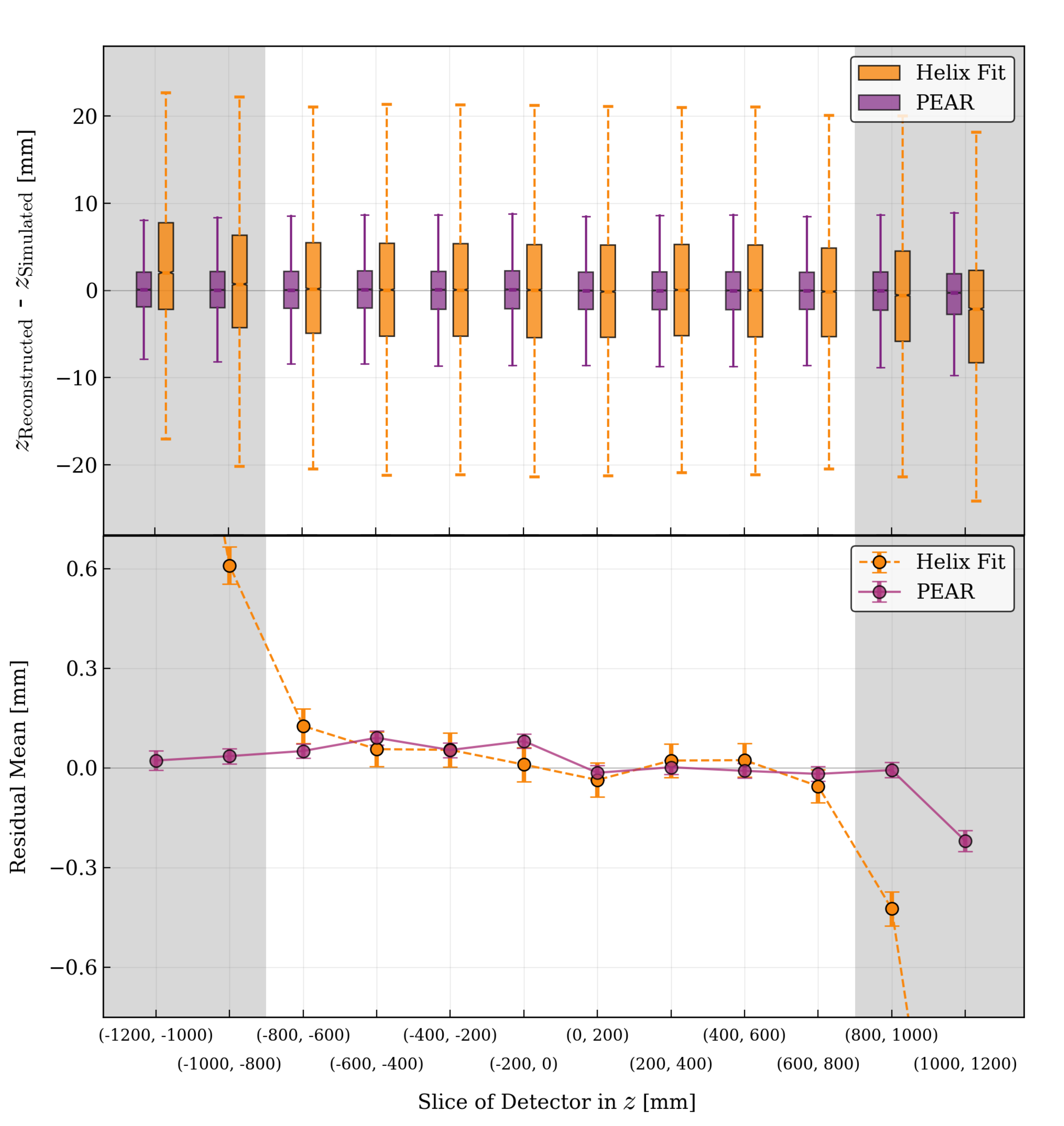}
\caption{\label{new_figs:box_and_bias}Box plot of residuals from PEAR and Helix Fit predictions for each 200\,mm slice of the detector (top). 
For this study, the line within the box denotes the median, the box covers the interquartile range (IQR), and the whiskers extend to the furthest residuals within 1.5 times the IQR on either side of the distribution. Residual mean ($\mu_i$ used in ARA to probe $z$-basis) for each 200\,mm slice of the detector (bottom). Note that the ends of the detector are greyed out, as generally no scientific measurements are made outside these bounds.}
\end{figure}

A box plot of the residuals is shown in Figure~\ref{new_figs:box_and_bias} (top), for slices of the detector along the $z$-axis, showing that PEAR has superior performance within each section of the detector. Figure~\ref{new_figs:box_and_bias} (bottom) shows the residual mean for each slice along the $z$-axis, and within the (-800,800)~mm region of interest, the absolute value of the residual mean is not larger than 0.11~mm for PEAR and 0.18~mm for Helix Fit. 
The observed increase in residual mean near the detector ends is due to reduced acceptance at the edges of the detector. 

\section{Conclusions and Outlook}

There are various additional optimization steps that could be undertaken to potentially further improve model performance. One such example is the use of more modern transformer-based architectures that have been shown to be highly capable of learning complex patterns from point clouds \cite{pointtransformerv3}.  
However, we decided to end optimization at this point, as the model achieved a performance that significantly surpassed the existing state-of-the-art, allowing us to prioritize other technical challenges. 

Owing to its novelty in antihydrogen physics, this paper does not address the issue of interpretability. While the deep learning process is not transparent, the results are in agreement with the expectations and provide an excellent starting point for future work. Additionally, unlike Helix Fit, this method currently only outputs the $z$ coordinate of the vertex. Work to extend this model to predict $x$ and $y$ coordinates of the vertex is currently underway and showing promising results. While this additional information is not needed for gravity measurements, it is useful for interpreting the physics results and to apply certain physics cuts to reduce the background (cosmic rays). Finally, it is desirable for the model to provide uncertainty estimates associated with its predictions. While this has yet to be implemented, we have noticed that high variance among individual PEAR model predictions within the ensemble is correlated with poor ensemble predictions and therefore plan to explore ensemble-based uncertainty estimation techniques.

Although PEAR has proven to be successful on simulated data, it must be validated on real data before use in the scientific analysis pipeline. If this performance transfers reasonably well to real data, then the method introduced in this paper will serve as an important tool in aiding ALPHA-g to carry out more precise measurements of the effect of gravity on antimatter and therefore help us better understand the fundamental building blocks of our universe. This validation is actively underway, and initial results show that PEAR works exceedingly well out of the box on calibration data.

This paper is solely meant to provide a brief summary of work on this project. As such, a more comprehensive paper that will provide a thorough description of the methods used here as well as additional validation of the model on simulation is currently in preparation. For now, more details can be found by following the links provided in Appendix~\ref{appendix:code}, particularly to the code repository where additional documentation is provided, and once available, a link to the full paper will be posted. 

\newpage
\begin{ack}
This work was supported by NSERC, NRC/TRIUMF, CFI. This research was enabled in part by the computational resources provided by the BC DRI Group and the Digital Research Alliance of Canada (alliancecan.ca) under Art Olin's allocations. The authors thank the anonymous reviewers who helped improve this manuscript as well as P. A. Amxaudruz, D. Bishop, M. Constable, P. Lu, L. Kurchaninov, K. Olchanski, F. Retiere, and B. Shaw (TRIUMF) for their work on the ALPHA-g detectors and the data acquisition system.

\end{ack}

\bibliographystyle{unsrtnat} 
\bibliography{biblio}

\begin{thebibliography}{8}
\providecommand{\natexlab}[1]{#1}
\providecommand{\url}[1]{\texttt{#1}}
\expandafter\ifx\csname urlstyle\endcsname\relax
  \providecommand{\doi}[1]{doi: #1}\else
  \providecommand{\doi}{doi: \begingroup \urlstyle{rm}\Url}\fi

\bibitem[So et~al.(2020)So, Fajans, and Bertsche]{Chukman2020}
Chukman So, Joel Fajans, and William Bertsche.
\newblock The alpha-g antihydrogen gravity magnet system.
\newblock \emph{IEEE Transactions on Applied Superconductivity}, 30\penalty0 (4):\penalty0 1--5, 2020.
\newblock \doi{10.1109/TASC.2020.2981272}.

\bibitem[Capra et~al.(2017)]{Capra2017}
Andrea Capra et~al.
\newblock {Design of a Radial TPC for Antihydrogen Gravity Measurement with ALPHA-g}.
\newblock \emph{JPS Conf. Proc.}, 18:\penalty0 011015, 2017.
\newblock \doi{10.7566/JPSCP.18.011015}.

\bibitem[CER(2021)]{CERN2021}
Alpha cools antimatter using laser light for the first time.
\newblock CERN Press Release, \url{https://home.cern/news/press-release/experiments/alpha-cools-antimatter-using-laser-light-first-time}, March 2021.
\newblock Accessed: 2024-09-08.

\bibitem[Anderson et~al.(2023)Anderson, Baker, Bonomi, Christensen, et~al.]{anderson2023observation}
EK~Anderson, CJ~Baker, G~Bonomi, A~Christensen, et~al.
\newblock Observation of the effect of gravity on the motion of antimatter.
\newblock \emph{Nature}, 621\penalty0 (7980):\penalty0 716--722, 2023.

\bibitem[Kuchera et~al.(2019)Kuchera, Ramanujan, Taylor, Strauss, Bazin, Bradt, and Chen]{Kuchera2019}
M.P. Kuchera, R.~Ramanujan, J.Z. Taylor, R.R. Strauss, D.~Bazin, J.~Bradt, and Ruiming Chen.
\newblock Machine learning methods for track classification in the at-tpc.
\newblock \emph{Nuclear Instruments and Methods in Physics Research Section A: Accelerators, Spectrometers, Detectors and Associated Equipment}, 940:\penalty0 156--167, 2019.
\newblock ISSN 0168-9002.
\newblock \doi{https://doi.org/10.1016/j.nima.2019.05.097}.
\newblock URL \url{https://www.sciencedirect.com/science/article/pii/S0168900219308046}.

\bibitem[Qi et~al.(2017)Qi, Su, Mo, and Guibas]{Qi2017}
Charles~R. Qi, Hao Su, Kaichun Mo, and Leonidas~J. Guibas.
\newblock Pointnet: Deep learning on point sets for 3d classification and segmentation.
\newblock In \emph{Proceedings of the IEEE Conference on Computer Vision and Pattern Recognition (CVPR)}, July 2017.

\bibitem[Breiman(1996)]{Breiman1996BaggingP}
L.~Breiman.
\newblock Bagging predictors.
\newblock \emph{Machine Learning}, 24:\penalty0 123--140, 1996.
\newblock URL \url{https://api.semanticscholar.org/CorpusID:47328136}.

\bibitem[Wu et~al.(2024)Wu, Jiang, Wang, Liu, Liu, Qiao, Ouyang, He, and Zhao]{pointtransformerv3}
Xiaoyang Wu, Li~Jiang, Peng-Shuai Wang, Zhijian Liu, Xihui Liu, Yu~Qiao, Wanli Ouyang, Tong He, and Hengshuang Zhao.
\newblock Point transformer v3: Simpler, faster, stronger.
\newblock In \emph{CVPR}, 2024.

\end{thebibliography}

\appendix
\section{Data and Code Availability}
\label{appendix:code}
To facilitate the reproducibility of our results and aid others interested in adapting our work, we have made our data and code publicly available, along with more documentation at each link below:
\begin{itemize}
    \item \textbf{Raw data} is generated with code available at \url{bitbucket.org/expalpha/alphasoft}.
    \item \textbf{Preprocessed data} is stored at \url{zenodo.org/records/13963779}.
    \item \textbf{Code} for all the data preparation steps, model training, and the full analysis for result generation, can be accessed at \url{gitlab.triumf.ca/alpha-ai/rTPC-AI}.
\end{itemize}

\section{Compute Resources}
\label{appendix:compute}
The training was performed using NVIDIA A100 GPUs on the Narval cluster managed by the Digital Research Alliance of Canada. The duration of training a single model was approximately 70 hours on a single A100 GPU with this number continuing to decrease as more GPU resources are utilized. Since we generated an ensemble of three models by selecting the best epoch from three different training runs, the total training time required on one GPU was therefore approximately 210 hours. Performing inference on a batch of 5,000 events with this ensemble takes only approximately 1 second to complete on the aforementioned GPU. These inference speeds are sufficiently fast for the ALPHA-g experiment's needs, given real-time processing is not necessary.

\end{document}